\documentclass[twocolumn,prl,aps,superscriptaddress,showpacs,amsmath,amssymb]{revtex4}
\usepackage{amsfonts}
\usepackage{bbm}

\usepackage{graphicx}
\usepackage{dcolumn}
\usepackage{bm}

\begin{document}

\title{Universal scheme to generate metal-insulator transition in disordered systems}
\author{Ai-Min Guo}
\affiliation{Institute of Physics, Chinese Academy of Sciences, Beijing 100190, China}
\author{Shi-Jie Xiong}
\affiliation{National Laboratory of Solid State Microstructures and Department of Physics, Nanjing University, Nanjing 210093, China}
\author{X. C. Xie}
\affiliation{International Center for Quantum Materials, Peking University, Beijing 100871, China}
\author{Qing-feng Sun}
\email{sunqf@iphy.ac.cn}
\affiliation{Institute of Physics, Chinese Academy of Sciences, Beijing 100190, China}

\begin{abstract}

We propose a scheme to generate metal-insulator transition in random binary layer (RBL) model, which is constructed by randomly assigning two types of layers. Based on a tight-binding Hamiltonian, the localization length is calculated for a variety of RBLs with different cross section geometries by using the transfer-matrix method. Both analytical and numerical results show that a band of extended states could appear in the RBLs and the systems behave as metals by properly tuning the model parameters, due to the existence of a completely ordered subband, leading to a metal-insulator transition in parameter space. Furthermore, the extended states are irrespective of the diagonal and off-diagonal disorder strengths. Our results can be generalized to two- and three-dimensional disordered systems with arbitrary layer structures, and may be realized in Bose-Einstein condensates.

\end{abstract}

\pacs{71.30.+h, 71.23.An, 72.20.Ee, 03.75.-b}

\maketitle

One of the most important issues in condensed-matter physics is Anderson localization \cite{apw1}, which predicts that the electronic wavefunctions may become localized in imperfect crystals and leads to a disorder-induced metal-insulator transition (MIT), owing to the quantum interference between multiple scatterings of an electron with random impurities and defects \cite{lpa1,kb1,la1}. Another seminal work along this direction is the scaling theory of localization \cite{ae1}, which indicates that all electronic states are exponentially localized in low-dimensional noninteracting systems even for infinitesimal disorder and become localized in three-dimensional (3D) systems with sufficiently large disorder strength. In fact, Anderson localization is a universal wave phenomenon and has been observed experimentally in a wide variety of systems, including light \cite{wds1,sm1,st1,ly1}, microwaves \cite{caa1}, acoustic waves \cite{hh1,fs1}, and matter waves \cite{bj1,rg1,spl1,kss1,jf1}.

\begin{figure}
\includegraphics[width=0.39\textwidth]{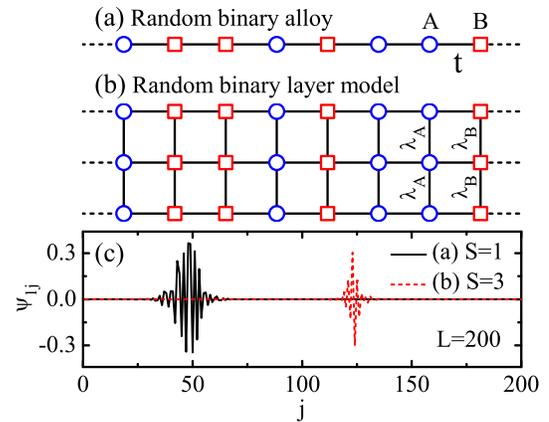}
\caption{\label{fig:one} (Color online) Schematic views of (a) random binary alloy and of (b) RBL model. The former is composed of two different sites (atoms) A and B with constant nearest-neighbor hopping integral $t$, while the latter is constituted by assigning two types of layers $L_A$ and $L_B$ at random, where $L_A$ ($L_B$) contains only A (B) sites. The coupling parameters within $L_A$ and $L_B$ layers are $\lambda_A$ and $\lambda_B$, respectively. (c) Typical wavefunctions of a random binary chain (solid line) and of a corresponding binary layer with $S=3$ (dashed line). The wavefunctions are obtained from direct diagonalization of the Hamiltonian with $\varepsilon _A$=$0$, $\varepsilon _B$=$2$, $t$=$\lambda _A$=$\lambda _B$=$1$, and $L$=$200$ [see Eq.~(\ref{eq1})]. All of the wavefunctions are bounded in a small region and both systems behave as insulators. However, a band of extended states will emerge in the RBLs by introducing a relation between $\varepsilon _A$, $\lambda _A$ and $\varepsilon _B$, $\lambda _B$ (see text).}
\end{figure}

The simplest one-dimensional (1D) example exhibiting Anderson localization is the random binary alloy, where the two constituents A and B are randomly distributed in the lattice [see Fig.~\ref{fig:one}(a)] and all eigenmodes are localized within a small region [see the solid line in Fig.~\ref{fig:one}(c)]. However, Anderson localization breaks down in the random binary alloy when correlations are introduced in the disorder distribution. A discrete number of extended states have been reported in the random-dimer model \cite{ddh1,whl1,pp1} and its generalized versions \cite{cx1,sjf1}, where one or both sites always appear in $n$-mer. A band of extended states will emerge in the binary alloy when the site energies are long-range correlated \cite{cp1}. In addition, other theoretical models have been suggested to produce conducting states in low-dimensional disordered systems \cite{dss1,dmf1,ra1,pa1,ss1,ggam1,bj2,bj3,bj4,gam1}, and some of them have been corroborated in GaAs-AlGaAs superlattices \cite{bv1} and Bose-Einstein condensate (BEC) \cite{rg1}. Nevertheless, we notice that all electronic states become localized in these correlated disordered systems when the disorder degree is extremely large.

In this Letter, we explore the electronic and localization properties of a random binary layer (RBL) model, which can be constructed by coupling identical random binary chains along the transverse direction or equivalently by stochastically assigning two types of layers $L_A$ and $L_B$ in the lattice, as illustrated in Fig.~\ref{fig:one}(b). One expects that since a single random binary chain behaves as an insulator, all electronic states should be localized in the RBL model [see the dashed line in Fig.~\ref{fig:one}(c)]. However, contrary to this physical intuition
and the scaling theory of localization, we show that a continuous band of extended states could emerge in the RBLs and the disordered systems behave as metals by introducing a relation for the model parameters, giving rise to an MIT in parameter space. This is due to the presence of a perfectly ordered subband. Besides, the mobility edges can be determined analytically and the extended states are independent of the strengths of the diagonal and off-diagonal disorders. Our results still hold in two-dimensional (2D) and 3D disordered systems with arbitrary layer structures, and may be implemented in the BECs.

The Hamiltonian of the RBL model can be written in the tight-binding form:
\begin{eqnarray}
{\cal H}=\sum_{i,j}\varepsilon_{j} c_{ij}^\dagger c_{ij}+ \sum_ {{\langle i,m\rangle},j} \lambda_j c_{ij}^\dagger c_{mj} + t \sum_ {i,{\langle j,n\rangle}} c_{ij}^\dagger c_{in}, \label{eq1}
\end{eqnarray}
where $c_{ij}^\dagger$ ($c_{ij}$) is the creation (annihilation) operator of an electron at site $(i,j)$, with subscripts $i \in [1,S]$ labeling a chain and $j \in [1,L]$ denoting a layer. $S$ is the number of chains and can also describe the cross section geometry, and $L$ is the length. The site energy $\varepsilon_{j}$ for the $L_A$ ($L_B$) layer is taken as $\varepsilon _A$ ($\varepsilon _B$), the interchain hopping integral $\lambda_j$ is set to $\lambda _A$ ($\lambda _B$) within the $L_A$ ($L_B$) layer, and $t$ is the intrachain hopping integral. $\langle \cdots \rangle$ represents the nearest-neighbor sites. In what follows, the most disordered case is considered with probability $\frac 1 2$ for each layer, $\varepsilon_A=0$ is chosen as the energy reference point, and all energy parameters are in units of $t$. Consequently, the diagonal disorder degree is $W= \varepsilon _B$.

It is convenient to study the localization properties of the RBLs by using the transfer-matrix method \cite{ma1,sk1}, i.e., the Lyapunov exponents $\gamma_i$'s can be obtained from the $\mathbf Q \mathbf R$ decomposition together with the Gram-Schmidt reorthonormalization. We calculate the first $S$ positive $\gamma_i$'s, which refer to the $S$ electron propagating modes along the longitudinal direction. The localization length $\xi$, measuring the extent of the wavefunction, is then defined as the reciprocal of the smallest positive $\gamma_i$. Our results are averaged over an ensemble of disorder configurations to reach convergence. Finally, the zero-temperature conductance is $G= (2e^2/h) \sum_ {i=1} ^S 1 /{\cosh^2 (\gamma_iL)}$ \cite{kb1,bcwj1}.

In the site representation, the Schr\"{o}dinger equation ${\cal H}| {\mathbf \Psi} \rangle= E|{\mathbf \Psi} \rangle$ can be expressed as
\begin{eqnarray}
(E{\mathbf I}- {\cal H}_j) {\mathbf \Psi}_j=t({\mathbf \Psi}_{j-1} + {\mathbf \Psi}_{j+1}). \label{eq2}
\end{eqnarray}
Here, $E$ is the Fermi energy, ${\mathbf I}$ is the $S\times S$ identity matrix, ${\cal H}_j$ is the sub-Hamiltonian matrix of the $j$th layer, and ${\mathbf \Psi}_j= (\psi_{1j},\psi_ {2j},\cdots, \psi_{Sj})^T$ with $\psi_{ij}$ the amplitude of the wavefunction at site $(i,j)$ and $T$ the transpose. It is clear that there are two different Hermitian matrices ${\cal H}_j$'s, namely ${\cal H}_A$ and ${\cal H}_B$. Both ${\cal H}_A$ and ${\cal H}_B$ can be diagonalized by a single unitary matrix $\mathbf U$ through $\mathbf P_j=\mathbf U^{\dagger} {\cal H}_j \mathbf U$, and the diagonal elements of $\mathbf P_j$ are the eigenvalues of ${\cal H}_j$. Thus, Eq.~(\ref{eq2}) can be transformed into
\begin{eqnarray}
(E{\mathbf I}- {\mathbf P}_j) {\mathbf \Phi}_j=t({\mathbf \Phi}_{j-1} + {\mathbf \Phi}_{j+1}),  \label{eq3}
\end{eqnarray}
with ${\mathbf \Phi}_j=\mathbf U^{\dagger} {\mathbf \Psi}_j= (\phi_{1j},\phi_{2j},\cdots, \phi_{Sj})^T$. Then the RBL model is decoupled into following Schr\"{o}dinger equations:
\begin{eqnarray}
(E- {\mathbf \nu}_{kj}) {\mathbf \phi}_{kj}=t({\mathbf \phi}_{kj-1} + {\mathbf \phi}_{kj+1}), \label{eq4}
\end{eqnarray}
where $\nu_{kj}$ is the $k$th eigenvalue of ${\cal H}_j$. When the cross section is a line, the eigenvalue $\nu_{kj}$ is
\begin{eqnarray}
\nu_{kj}= \varepsilon_j+ 2\lambda_j \cos \frac {k\pi} {S+1};  \label{eq5}
\end{eqnarray}
when the cross section is a rectangle with $S$=$S_y \times S_z$, the $\nu_{kj}$ with integer index $k$=$(k_y-1)S_z+k_z$ is
\begin{eqnarray}
\nu_{kj}= \varepsilon_j+ 2\lambda_j (\cos \frac {k_y\pi}  {S_y+1} + \cos \frac {k_z \pi} {S_z+1}).  \label{eq6}
\end{eqnarray}
Here, $\varepsilon _j $$=$$0$ and $\lambda _j$$=$$\lambda _A$ when ${\cal H}_j$$=$${\cal H}_A$, $\varepsilon _j$$=$$ \varepsilon _B$ and $\lambda _j$$=$$\lambda _B$ when ${\cal H}_j $$=$${\cal H}_B$, $k\in[1,S]$, $k_y \in [1, S_y]$, and $k_z \in [1, S_z]$. It clearly appears that the RBLs can be decoupled into $S$ chains, each of which is a random binary chain with site energies $\nu_{kA}$ and $\nu_{kB}$ determined by Eqs.~(\ref{eq5}) or (\ref{eq6}); and the electronic and localization properties of the RBLs could be strongly changed by varying $\lambda _j$ only, and the energy spectrum will be shifted. One notices that when $\nu_{kA} =\nu_{kB}$, the $k$th decoupled chain is a perfectly ordered one with constant site energy $\nu _{kA}$ and nearest-neighbor hopping integral $t$, and the electronic states are Bloch-like states in the energy interval $[\nu_{kA}-2t, \nu_{kA}+2t]$, with the mobility edges being at $E= \nu_{kA}\pm 2t$. This result will be verified numerically for several RBLs with different cross sections (see below).

\begin{figure}
\includegraphics[width=0.39\textwidth]{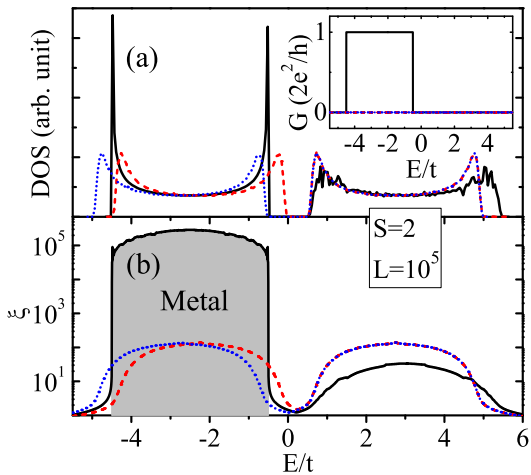}
\caption{\label{fig:two}(Color online) (a) DOS and (b) localization length for disordered two-leg ladders with $\varepsilon_ B$=$0.5$, $\lambda_ A$=$-2.5$, $\lambda_ B$=$-3$ (solid lines), $\varepsilon_ B$=$0.5$, $\lambda_A$=$\lambda_ B$=$-2.5$ (dashed lines), and $\varepsilon_ B$=$0$, $\lambda_ A$=$-2.5$, $\lambda_ B$=$-3$ (dotted lines). The inset displays the corresponding conductance \cite{note1}. The DOS is calculated by diagonalizing the Hamiltonian for a single disorder configuration and is almost the same when other disorder configurations are employed, since the system size is very large. In the gray energy region, $\xi$ is averaged in a small energy window of width 0.1 around $E$ to avoid numerical fluctuations.}
\end{figure}

\begin{figure}
\includegraphics[width=0.46\textwidth]{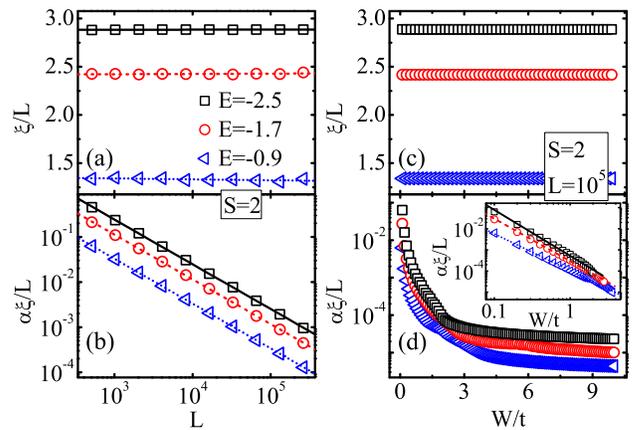}
\caption{\label{fig:three}(Color online) Scaling behaviors of the normalized localization length at several energies for the corresponding two-leg ladders in Fig.~\ref{fig:two}. The states are extended in the thermodynamic limit if $\frac \xi L$ is invariant or increased by increasing $L$ and are localized otherwise. $\frac \xi L$ vs $L$ for (a) $\varepsilon_ B$=$0.5$, $\lambda_ A$=$-2.5$, $\lambda_ B$=$-3$ and for (b) $\varepsilon_ B$=$0.5$, $\lambda_A$=$\lambda_ B$=$-2.5$. The dependence of $\frac \xi L$ on $L$ can be fitted exactly by a simple function ${\frac \xi L} \varpropto L^ \beta$, as shown by the solid, dashed, and dotted lines. $\frac \xi L$ vs $W$ $(\varepsilon_B)$ for (c) $\lambda_ A$=$-2.5$, $\lambda_ B$=$-2.5-W$ and for (d) $\lambda_A$=$\lambda_ B$=$-2.5$. The inset presents the enlarged view of $\frac \xi L$-$W$ in the region of $W<W_c$ with $W_c$=$E+2t-\lambda_B$, and the corresponding fitting curves $\frac \xi L \varpropto W^ \beta$. The parameter $\alpha$ in the vertical scale of (b) and (d) is employed to separate the lines. $\alpha $=$2$, $1$, and $0.5$ for $E$=$-2.5$, $-1.7$, and $-0.9$, respectively.}
\end{figure}

We first consider the simplest case of $S$=$2$, namely the two-leg ladder model \cite{gam1}, which has been used to simulate the charge transport through double-stranded DNA. In this respect, the site energies of two decoupled chains are $\nu _{1A}$=$\lambda _A$, $\nu _{1B} $=$ \varepsilon _B+ \lambda _B$ and $\nu _{2A}$=$-\lambda _A$, $\nu _{2B}$=$ \varepsilon _B-\lambda _B$, respectively. Accordingly, the energy spectrum of the disordered two-leg ladder, composed of the energy bands of individual decoupled chains, can be written as
\begin{equation}
\begin{split}
[\nu _{1A}-2t,\nu _{1A}+2t]\cup[\nu _{1B}-2t,\nu _{1B} +2t], \\
[\nu _{2A}-2t,\nu _{2A}+2t]\cup[\nu _{2B}-2t,\nu _{2B} +2t]. \label{eq7}
\end{split}
\end{equation}
Figures~\ref{fig:two}(a) and \ref{fig:two}(b) display respectively the density of states (DOS) and the localization length $\xi$ with $\varepsilon_ B$=$0.5$, $\lambda_ A$=$-2.5$, $\lambda_ B$=$-3$ (solid lines), so that $\nu _{1A}=\nu _{1B}$. One can see from Fig.~\ref{fig:two}(a) that the energy spectrum consists of two subbands divided by an energy gap. The left subband resembles a periodic chain with extremely sharp van Hove singularities and smooth DOS profile, although both diagonal and off-diagonal disorders exist in the system; while the right subband refers to a disordered chain. This accords with Eq.~(\ref{eq7}), indicating that each subband corresponds to an individual decoupled chain. On the other hand, it is clear that $\xi$ is comparable to $L$ for the left subband [Fig.~\ref{fig:two}(b)] and the conductance $G$ is exactly ${2e ^2}/ h$ (see inset), implying a perfect propagating mode in this energy region. Actually, all states in the left subband are extended in the thermodynamic limit, as verified by the scaling behaviors of the normalized localization length $\frac \xi L$ in Fig.~\ref{fig:three}(a). We find that the scaling function $\beta$=$0$ for all energies in the left subband \cite{ae1} and the disordered two-leg ladder exhibits metallic behavior. The states are localized elsewhere with $G$=$0$ and the mobility edges are at $E=\lambda_{A}\pm2t$. In fact, these results always hold when $\nu _{1A}= \nu _{1B}$ or $\nu _{2A}= \nu _{2B}$ is satisfied, regardless of the diagonal and off-diagonal disorder degrees and the other parameters. Besides, the position of the energy band of extended states, relative to that of localized states, can be modulated by altering the model parameters [Eq.~(\ref{eq7})] and a variety of phenomena will occur, such as the shift of the mobility edges. These statements will be further discussed in the following.

As a comparison, Fig.~\ref{fig:two} shows DOS and $\xi$ by changing only one parameter of the two-leg ladder, viz., replacing $\lambda_ B$ with $-2.5$ (dashed lines) and $\varepsilon_ B$ with $0$ (dotted lines), respectively. In contrast, all states become localized for both two-leg ladders with $G=0$ (see inset) and constant localization length at fixed energies [from Fig.~\ref{fig:three}(b) we obtain $\beta$=$-1$], due to the Anderson localization effects, although they possess either the diagonal disorder (dashed lines) or the off-diagonal disorder (dotted lines) and are more ordered than the former case.

Figures~\ref{fig:three}(c) and \ref{fig:three}(d) plot $\frac \xi L$ vs $W$ ($\varepsilon _B$) for the two-leg ladders with $\lambda _A$=$-2.5$, $\lambda _B$=$-2.5-W$ and $\lambda _A$=$\lambda _B$=$-2.5$, respectively. Contrary to the previous works that all states become localized in the disordered systems when the disorder degree is very large \cite{ddh1,whl1,pp1,cx1,sjf1,cp1,dss1,dmf1,ra1,pa1,ss1,ggam1,bj2,bj3,bj4,gam1}, we can see from Fig.~\ref{fig:three}(c) that $\frac \xi L$ is independent of $W$ for the former ladder and all states are always extended in the gray energy region [Fig.~\ref{fig:two}(b)], because $\nu _{1A}=\nu _{1B}$=$-2.5$ and the left subband denotes a completely ordered chain and does not move with $W$. While for the latter ladder, a crossover $W_c$, dividing strong and weak dependence of $\frac \xi L$ on $W$, exists in all the curves [Fig.~\ref{fig:three}(d)], similar to that observed in the experiments \cite{st1,bj1,jf1}. $\frac \xi L$ is strongly declined by increasing $W$ up to $W_c$=$E+2t-\lambda_B$, due to the gradually enhanced scatterings from the potential barriers at B sites. The behavior of $\frac \xi L$ vs $W$ can be fitted by a power law relation for $W<W_c$ (see inset). $\frac \xi L$ is slightly decreased by further increasing $W$, since the energy $E$ is separated from the band $[\nu _{1B}-2t, \nu _{1B}+2t]$ for $W>W_c$ [see Eq.~(\ref{eq7})]. In this situation, the state at $E$ is formed by A sites only and the electron is usually confined inside a single A site, leading to the localization length $\xi \approx1$.

\begin{figure}
\includegraphics[width=0.42\textwidth]{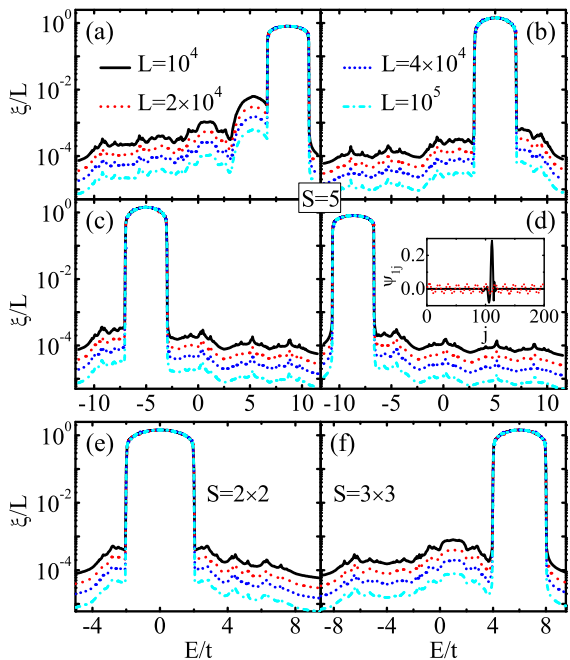}
\caption{\label{fig:four}(Color online) Energy-dependent normalized localization length for the RBLs of linear and square cross sections with several values of $L$. $\frac \xi L$ vs $E$ for (a) $\varepsilon_ B$=$\sqrt3$, $\lambda_ B$=$4$, for (b) $\varepsilon_ B$=$3$, $\lambda_ B$=$2$, for (c) $\varepsilon_ B$=$-4$, $\lambda_ B$=$1$, and for (d) $\varepsilon_ B$=$-5\sqrt3$, $\lambda_ B$=$0$ with $S$=$5$ and $\lambda _A$=$5$. $\frac \xi L$ vs $E$ for (e) $S$=$2$$\times$$2$, $\varepsilon_ B$=$2$, $\lambda_ A$=$3$, $\lambda_ B$=$2$ and for (f) $S$=$3$$\times$$3$, $\varepsilon_ B$=$2$, $\lambda_ A$=$0$, $\lambda_ B$=$\sqrt 2$. The inset presents the wavefunctions of two eigenmodes at $E$=$-6.9901$ (solid line) and $E$=$-6.8110$ (dashed line) in the metallic regime with $L$=$200$. The arrangement of A and B sites is the same as that in Fig.~\ref{fig:one}(c).}
\end{figure}

We then consider the RBLs of various cross sections. Figures~\ref{fig:four}(a)-\ref{fig:four}(d) show $\frac \xi L$ vs $E$ for several values of $\varepsilon _B$, $\lambda _B$, $L$ by fixing $S$=$5$ and $\lambda _A$=$5$. The $\varepsilon _B$ and $\lambda _B$ are chosen so that $\nu _{kA}=\nu _{kB}$ is satisfied for specific $k$ and a perfectly ordered chain always appears. A band of extended states as well as the mobility edges can be observed in all the situations and are shifted toward lower energies by increasing $W$; and the wavefunction in the metallic regime will spread over the entire system (see the dashed line of the inset). In this regard, the RBLs may be used as switching devices and their electronic structure could be controlled by changing $\varepsilon _B$ and $\lambda _B$ while keeping other parameters fixed. However, the localized state can survive in the metallic regime (see the solid line of the inset), due to the superposition of extended and localized subbands, as can be inferred from Eq.~(\ref{eq5}). Similar results can be obtained for the RBLs of square cross sections, even when $\lambda _A$=$0$ [Figs.~\ref{fig:four}(e)-\ref{fig:four}(f)]. Therefore, we conclude that a band of extended states will appear in the RBLs and the systems behave as metals by implementing a relation between $\varepsilon _A$, $\lambda _A$ and $\varepsilon _B$, $\lambda _B$, leading to an MIT in parameter space. These are irrespective of the diagonal and off-diagonal disorder strengths.

Finally, we discuss the generalization of our results and the possible relevance to realistic systems. (i) In contrast to the random-dimer model that the extended states vanish when it contains more than two different sites \cite{ddh1,whl1,pp1,cx1,sjf1}, our results can still hold when the site energies are randomly selected from a certain probability distribution. For example, we consider $\varepsilon _j$ uniformly distributed within $[-\frac W 2, \frac W 2]$. By properly choosing the interchain hopping integral that $\nu_{kj} $=$E_c$ always holds for specific $k$ [Eq.~(\ref{eq5})] or $k_y$, $k_z$ [Eq.~(\ref{eq6})] with $E_c$ a constant, a band of extended states will emerge in the energy region $[E_c-2t, E_c+2t]$. (ii) Our results remain valid for other 2D and 3D disordered systems when $\nu_{kj} $=$E_c$ is satisfied. For an $S$$\times$$ L$ disordered system, the number of the condition $\nu_{kj} $=$E_c$ is $\frac {2S-1+(-1)^S} 2$, except for the case $\cos \frac {k\pi} {S+1}$=$0$ which leads to an ordered system. Similarly for an $S$$\times$$ S$$\times L$ disordered system, the number is $\frac {2S^2-1 +(-1)^S} 4$. (iii) These results also hold for the disordered systems with various layer structures, such as rectangle and hexagon, as long as the $\nu_{kj}$'s for fixed $k$ (or $k_y$, $k_z$) are constant or follow a periodic distribution  by increasing $j$. (iv) Besides the double-stranded DNA, we believe our results to be relevant for the BECs, which offer an ideal platform for studying Anderson localization \cite{bj1,rg1,spl1,kss1,jf1}; and the model parameters can be precisely controlled \cite{rg2}.

In summary, we suggest a scheme to produce metallic states in a series of disordered systems. Our results indicate that a band of extended states could emerge in the thermodynamic limit and the disordered systems behave as metals by properly adjusting the model parameters, generating a metal-insulator transition in parameter space. The results are independent of the diagonal and off-diagonal disorder strengths, and still hold in two- and three-dimensional disordered systems.

\section*{Acknowledgments}

This work was financially supported by NBRP of China (2012CB921303 and 2009CB929100) and NSF-China under Grants Nos. 10974236 and 11074174.


\begin{references}

\bibitem{apw1}  P. W. Anderson, Phys. Rev. {\bf 109}, 1492 (1958).
\bibitem{lpa1}  P. A. Lee and T. V. Ramakrishnan, Rev. Mod. Phys. {\bf 57}, 287 (1985).
\bibitem{kb1}   B. Kramer and A. MacKinnon, Rep. Prog. Phys. {\bf 56}, 1469 (1993).
\bibitem{la1}   A. Lagendijk, B. van Tiggelen, and D. S. Wiersma, Phys. Today {\bf 62}, 24 (2009).
\bibitem{ae1}   E. Abrahams, P. W. Anderson, D. C. Licciardello, and T. V. Ramakrishnan, Phys. Rev. Lett. {\bf 42}, 673 (1979).
\bibitem{wds1}  D. S. Wiersma, P. Bartolini, A. Lagendijk, and R. Righini, Nature (London) {\bf 390}, 671 (1997).
\bibitem{sm1}   M. St\"{o}rzer, P. Gross, C. M. Aegerter, and G. Maret, Phys. Rev. Lett. {\bf 96}, 063904 (2006).
\bibitem{st1}   T. Schwartz, G. Bartal, S. Fishman, and M. Segev, Nature (London) {\bf 446}, 52 (2007).
\bibitem{ly1}   Y. Lahini, A. Avidan, F. Pozzi, M. Sorel, R. Morandotti, D. N. Christodoulides, and Y. Silberberg, Phys. Rev. Lett. {\bf 100}, 013906 (2008).
\bibitem{caa1}  A. A. Chabanov, M. Stoytchev, and A. Z. Genack, Nature (London) {\bf 404}, 850 (2000).
\bibitem{hh1}   H. Hu, A. Strybulevych, J. H. Page, S. E. Skipetrov, and B. A. van Tiggelen, Nature Phys. {\bf 4}, 945 (2008).
\bibitem{fs1}   S. Faez, A. Strybulevych, J. H. Page, A. Lagendijk, and B. A. van Tiggelen, Phys. Rev. Lett. {\bf 103}, 155703 (2009).
\bibitem{bj1}   J. Billy, V. Josse, Z. Zuo, A. Bernard, B. Hambrecht, P. Lugan, D. Cl\'{e}ment, L. Sanchez-Palencia, P. Bouyer, and A. Aspect, Nature (London) {\bf 453}, 891 (2008).
\bibitem{rg1}   G. Roati, C. D'Errico, L. Fallani, M. Fattori, C. Fort, M. Zaccanti, G. Modugno, M. Modugno, and M. Inguscio, Nature (London) {\bf 453}, 895 (2008).
\bibitem{spl1}  L. Sanchez-Palencia and M. Lewenstein, Nature Phys. {\bf 6}, 87 (2010).
\bibitem{kss1}  S. S. Kondov, W. R. McGehee, J. J. Zirbel, and B. DeMarco, Science {\bf 334}, 66 (2011).
\bibitem{jf1}   F. Jendrzejewski, A. Bernard, K. M\"{u}ller, P. Cheinet, V. Josse, M. Piraud, L. Pezz\'{e}, L. Sanchez-Palencia, A. Aspect, and P. Bouyer, Nature Phys. {\bf 8}, 398 (2012).
\bibitem{ddh1}  D. H. Dunlap, H.-L. Wu, and P. W. Phillips, Phys. Rev. Lett. {\bf 65}, 88 (1990).
\bibitem{whl1}  H.-L. Wu and P. Phillips, Phys. Rev. Lett. {\bf 66}, 1366 (1991).
\bibitem{pp1}   P. Phillips and H.-L. Wu, Science {\bf 252}, 1805 (1991).
\bibitem{cx1}   X. Chen and S. Xiong, Phys. Lett. A {\bf 179}, 217 (1993).
\bibitem{sjf1}  J.-F. Schaff, Z. Akdeniz, and P. Vignolo, Phys. Rev. A {\bf 81}, 041604(R) (2010).
\bibitem{cp1}   P. Carpena, P. Bernaola-Galv\'{a}n, P. C. Ivanov, and H. E. Stanley, Nature (London) {\bf 418}, 955 (2002).
\bibitem{dss1}  S. Das Sarma, S. He, and X. C. Xie, Phys. Rev. Lett. {\bf 61}, 2144 (1988).
\bibitem{dmf1}  F. A. B. F. de Moura and M. L. Lyra, Phys. Rev. Lett. {\bf 81}, 3735 (1998).
\bibitem{ra1}   A. Rodr\'{i}guez, V. A. Malyshev, G. Sierra, M. A. Mart\'{i}n-Delgado, J. Rodr\'{i}guez-Laguna, and F. Dom\'{i}nguez-Adame, Phys. Rev. Lett. {\bf 90}, 027404 (2003).
\bibitem{pa1}   A. Punnoose and A. M. Finkel'stein, Science {\bf 310}, 289 (2005).
\bibitem{ss1}   S. Sil, S. K. Maiti, and A. Chakrabarti, Phys. Rev. Lett. {\bf 101}, 076803 (2008); Phys. Rev. B {\bf 78}, 113103 (2008).
\bibitem{ggam1} A. M. Garc\'{i}a-Garc\'{i}a and E. Cuevas, Phys. Rev. B {\bf 79}, 073104 (2009).
\bibitem{bj2}   J. Biddle, B. Wang, D. J. Priour Jr., and S. Das Sarma, Phys. Rev. A {\bf 80}, 021603(R) (2009).
\bibitem{bj3}   J. Biddle and S. Das Sarma, Phys. Rev. Lett. {\bf 104}, 070601 (2010).
\bibitem{bj4}   J. Biddle, D. J. Priour Jr., B. Wang, and S. Das Sarma, Phys. Rev. B {\bf 83}, 075105 (2011).
\bibitem{gam1}  A.-M. Guo and S.-J. Xiong, Phys. Rev. B {\bf 83}, 245108 (2011).
\bibitem{bv1}   V. Bellani, E. Diez, R. Hey, L. Toni, L. Tarricone, G. B. Parravicini, F. Dom\'{i}nguez-Adame, and R. G\'{o}mez-Alcal\'{a}, Phys. Rev. Lett. {\bf 82}, 2159 (1999).
\bibitem{ma1}   A. MacKinnon and B. Kramer, Phys. Rev. Lett. {\bf 47}, 1546 (1981); Z. Phys. B {\bf 53}, 1 (1983).
\bibitem{sk1}   K. Slevin, Y. Asada, and L. I. Deych, Phys. Rev. B {\bf 70}, 054201 (2004).
\bibitem{bcwj1} C. W. J. Beenakker, Rev. Mod. Phys. {\bf 69}, 731 (1997).
\bibitem{note1} During the final step of the $\mathbf {QR}$ decomposition, the contribution of the orthogonal matrix to $\gamma_i$'s is considered, see Ref.~\cite{sk1} for details.
\bibitem{rg2}   G. Roati, M. Zaccanti, C. D'Errico, J. Catani, M. Modugno, A. Simoni, M. Inguscio, and G. Modugno, Phys. Rev. Lett. {\bf 99}, 010403 (2007) and references therein.

\end{references}
\end{document}